______________________________________________________________________________

\documentstyle[12pt]{article}
\begin{document}

\title{Algebraic description \\
of anharmonic stretching vibrations }

\author{Xi-Wen Hou$^{1,2}$, Shi-Hai Dong$^1$,  Zong-Liang Fang$^1$,
and Zhong-Qi Ma$^1$ \\
{\footnotesize $^1$ Institute of High Energy Physics, 
P.O. Box 918, Beijing 100039, People's Republic of China} \\
{\footnotesize $^2$ Department of Physics, University of Three Gorges, 
Yichang 443000, People's Republic of China }}

\date{}

\maketitle

\vspace{6mm}

\begin{abstract}
A Fermi resonance-algebraic model is proposed for molecular vibrations, 
where a U(2) algebra is used for describing the vibrations of each bond,
and Fermi resonances between stretching and bending modes are taken 
into account. The model for a bent molecule XY$_2$ and a molecule XY$_3$
is successfully applied to fit the recently
observed vibrational spectrum of the water molecule and arsine (AsH$_3$), 
respectively, and results are compared with those of other models.
Calculations show that algebraic approaches can be used as an effective 
method for describing molecular vibrations with small standard deviations.

\vspace{3mm}

Keywords~~~~~U(2) algebra~~~~~Vibrational spectra~~~~~Fermi resonance

\end{abstract}

\newpage
Algebraic methods have been applied to molecular systems
for a number of years, and a recent review $^{[1]}$ is available. 
U(4) and U(2) algebraic model have mostly been used so far for a 
description of rovibrational spectra of molecules. U(4) model has 
two advantages that rotations and vibrations are treated simultaneously, 
and that Fermi resonances are described by the nondiagonal matrix elements of 
Majorana operators. But it becomes quite complicated when the number of atoms 
in a molecule is larger than 4. U(2) model is particularly well suited 
for dealing with the stretching vibrations of polyatomic molecules, 
and currently under further development $^{[2\sim 5]}$. 

Recently, an important development on these methods is to incorporate 
bending vibrational modes of polyatomic molecules. Sample molecules 
such as acetylene $^{[6]}$, as well as X$_3$ $^{[2]}$ and XY$_4$ $^{[5,7]}$
molecules have been studied via algebraic methods. These approaches 
suffer from the neglect of the Fermi resonance interactions between 
stretching and bending modes. Due to their importance as a mechanism 
for intramolecular energy transfer and for descriptions of excited 
stretching and bending vibrations, Fermi resonances have been taken 
into account in simple Fermi resonance-local mode models for bent 
triatomic $^{[8]}$ and pyramidal XH$_3$ $^{[9]}$ systems and in 
boson-realization models for bent triatomic $^{[10]}$ and XY$_4$ $^{[11]}$ 
molecules. In this paper we will introduce a Fermi resonance-algebraic 
model, where a U(2) algebra is used for describing the vibration 
of each bond. Our method is addressed for a bent molecule XY$_2$ 
and a molecule XY$_3$, where Fermi resonances between the 
stretch and the bend are considered. In a limit, the model corresponds 
to the boson-realization model. 
Vibrational spectra of two sample molecules H$_2$O and AsH$_3$  
are used to test our model. This method can be extended for 
describing the vibrations of molecules with other symmetry.
The corresponding results will be discussed elsewhere.

\vspace{5mm}

\begin{center}
{\large \bf 1. Hamiltonian}
\end{center}

Let $n$ be the total degrees of freedom of vibrations and $m$ 
the degrees of freedom of stretching vibrations in a polyatomic molecule.
We introduce $n$ U(2) algebras to describe vibrations: U$_i$(2) 
($1\leq i \leq m$) for stretching modes and U$_{\mu}$(2) 
($1+m\leq \mu \leq n$) for bending modes. 
Each U$_{\alpha}$(2) ($1\leq \alpha \leq n$) is generated by the 
operators $\{ \hat{N}_{\alpha},\hat{J}_{+,\alpha} \hat{J}_{-,\alpha}, 
\hat{J}_{0,\alpha} \}$, satisfying the following commutation relations:  
$$\begin{array}{ll}
~[\hat{J}_{0,\alpha}, \hat{J}_{\pm,\beta}]=\pm \delta_{\alpha \beta}
\hat{J}_{\pm,\alpha},~~~
&[\hat{J}_{+,\alpha}, \hat{J}_{-,\beta}]=2 \delta_{\alpha \beta} 
\hat{J}_{0,\alpha},\\
~[\hat{N}_{\alpha}, \hat{J}_{0,\beta}]=0,
~~&[\hat{N}_{\alpha}, \hat{J}_{\pm,\beta}]=0 .
\end{array}  \eqno (1.1)  $$

\noindent
where $\hat{N}_{\alpha}$ is related with the Casimir operator
of U(2):
$$2\hat{J}_{0,\alpha}^{2}+\hat{J}_{+,\alpha}\hat{J}_{-,\alpha}
+\hat{J}_{-,\alpha}\hat{J}_{+,\alpha}
=\hat{N}_{\alpha}(\hat{N}_{\alpha}/2+1). $$

The local basis states for each bond are written as 
$|N_{\alpha},v_{\alpha}\rangle$, where $N_{\alpha}$ is the eigenvalue 
of $\hat{N}_{\alpha}$ and $v_{\alpha}$ the number of quanta on the 
$\alpha$th bond. The action of $\hat{J}_{\pm,\alpha}$ 
on the local states is given by
$$\begin{array}{rl}
~\hat{J}_{+,\alpha}~|N_{\alpha},v_{\alpha}\rangle &=
\sqrt{v_{\alpha}(N_{\alpha}-v_{\alpha}+1)}~|N_{\alpha},v_{\alpha}-1\rangle,\\
~\hat{J}_{-,\alpha}~|N_{\alpha},v_{\alpha}\rangle &=
\sqrt{(v_{\alpha}+1)(N_{\alpha}-v_{\alpha})}~|N_{\alpha},v_{\alpha}+1\rangle.
\end{array}  \eqno (1.2) $$

\noindent
Those $N_{\alpha}$ of equivalent bonds are equal to each other: $N_{i}= N_s$, 
and $N_{\mu}= N_b$, where footnotes $s$ and $b$ refer to the stretching 
and the bending vibrations, respectively.

Due to Fermi resonances in the considered system, we will take 
$V~=~V_s~+~V_b~/2$ as a conservative quantity, 
where $V_s$ and $V_b$ is the total number of phonons
of stretching and bending modes, respectively.
Define the scale transformations for convenience:
$$~a_{\alpha}~\equiv~\hat{J}_{+,\alpha}~/\sqrt{N_{\alpha}},~~~~
~a^{\dag}_{\alpha}~\equiv~\hat{J}_{-,\alpha}~/\sqrt{N_{\alpha}}.
\eqno (1.3) $$

\noindent
In the limit of $N_s \rightarrow \infty$ and $N_b \rightarrow \infty$,
$a_{\alpha}(a^{\dag}_{\alpha})$ reduce to bosonic operators with usual
boson commutation relations $^{[2]}$.

In the following, we will present the vibrational model in terms of 
Eq. (1.3) for two molecular systems with different symmetries.

\vspace{2mm}
\noindent
{\bf 1.1 The model for vibrations of a bent molecule XY$_2$}

\vspace{2mm}
For a bent triatomic molecule, we have $n=3$ and $m=2$. The Hamiltonian, 
where the interactions are taken up to the fifth orders, is given as 
follows:

$$\begin{array}{rl}
H&=~\displaystyle  \sum_{i=1}^{2} a_{i}^{\dagger}a_{i}
(\omega_{s}+x_{s}a_{i}a_{i}^{\dagger}) +a_3^{\dagger}a_3
(\omega_{b}+x_{b}a_3a_3^{\dagger})+\lambda_{1} 
(a_{1}^{\dagger}a_{2}+a_{2}^{\dagger}a_{1}) \\
&~~~+~\lambda_{2} (a_{1}^{\dagger}a_{1}a_{2}^{\dagger}a_{2})
+\lambda_{3} (a_{1}^{\dagger}a_{1}^{\dagger}a_{2}a_{2}+H.c.)
+\lambda_{4} (a_{1}^{\dagger}a_{2}+a_{2}^{\dagger}a_{1})
(a_{1}^{\dagger}a_{1}+a_{2}^{\dagger}a_{2}) \\
&~~~+~\lambda_{5}(a_{1}^{\dagger}a_{2}+a_{2}^{\dagger}a_{1})
a_3^{\dagger}a_3 
+\lambda_{6}(a_{1}^{\dagger}a_{1}+a_{2}^{\dagger}a_{2})
a_3^{\dagger}a_3
+\lambda_{7}\{(a_{1}^{\dagger}+a_{2}^{\dagger})a_3a_3
+H.c.\} \\
&~~~+~\lambda_{8}\{a_{1}^{\dagger}a_{2}^{\dagger}(a_{1}+a_{2})
a_{3}a_{3}+H.c.\}
~+~\lambda_{9}\{(a_{1}^{\dagger}a_{1}^{\dagger}a_{1}+
a_{2}^{\dagger}a_{2}^{\dagger}a_{2})a_3a_3+H.c.\}\\
&~~~+~\lambda_{10}\{(a_{1}^{\dagger}a_{1}^{\dagger}a_{2}
+a_{2}^{\dagger}a_{2}^{\dagger}a_{1})a_3a_3+H.c.\}
+\lambda_{11}\{(a_{1}^{\dagger}+a_{2}^{\dagger})
a_3^{\dagger}a_3a_{3}a_{3}+H.c.\},
\end{array} \eqno (1.4) $$

\noindent                                   
where $\omega$ and $x$ are used for comparison with
the constants in the Morse potential.
$\lambda_{i}$ $(1\leq i \leq 11)$ are the coupling constants. 
The term with $\lambda_7$ describes the Fermi resonance between the 
stretching and the bending vibrations. The last four terms in Eq. (1.4) 
are the fifth order interactions caused by Fermi resonance.   
In the limit of N$_s$ $\rightarrow \infty$ and N$_b$ 
$\rightarrow \infty$, Eq. (1.4) corresponds to the boson-realization
model $^{[10]}$.

\vspace{2mm}
\noindent
{\bf 1.2 The model for vibrations of a molecule XY$_3$}

\vspace{2mm}
For a $C_{3v}$ symmetric molecule, we have $n=6$ and $m=3$. 
From the standard method of group theory, it is not difficult to obtain
the character table, the representation matrices of the generator,
and the Clebsch-Gordan coefficients of the $C_{3v}$ group. From that
knowledge, Hamiltonian where the interactions are taken up to the 
fourth orders are expressed in terms of operators Eq. (1.3):
$$\begin{array}{rl}
H&=~\displaystyle \sum_{i=1}^{3}~ a_{i}^{\dagger}a_{i}
(\omega_{s}~+~x_{s}~a_{i}a_{i}^{\dagger}) 
~+~\displaystyle \sum_{\mu=4}^{6}~ a_{\mu}^{\dagger}a_{\mu}
(\omega_{b}~+~x_{b}~a_{\mu}a_{\mu}^{\dagger})  
~+~\eta_{1}~\displaystyle \sum_{i \neq j}~ a_{i}^{\dagger}a_{j} \\
&~~~+~\eta_{2}~\displaystyle \sum_{\mu \neq \nu}~ a_{\mu}^{\dagger}a_{\nu}
~+~\eta_{3}~(a_{1}^{\dagger}a_{4}a_{4}~+~a_{2}^{\dagger}a_{5}a_{5}
~+~a_{3}^{\dagger}a_{6}a_{6}~+~H.c.) \\
&~~~+~\eta_{4}~\{a_{1}^{\dagger}(a_{5}a_{5}~+~a_{6}a_{6})~+~
a_{2}^{\dagger}(a_{4}a_{4}~+~a_{6}a_{6})~+~a_{3}^{\dagger}(a_{4}a_{4}~
+~a_{5}a_{5})~+~H.c.\}   \\
&~~~+~\eta_{5}~\{a_{1}^{\dagger}a_{4}(a_{5}~+~a_{6})~+~
a_{2}^{\dagger}a_{5}(a_{4}~+~a_{6})~+~
a_{3}^{\dagger}a_{6}(a_{4}~+~a_{5})~+~H.c.\} \\
&~~~+~\eta_{6}~(a_{1}^{\dagger}a_{5}a_{6}~+~a_{2}^{\dagger}a_{4}a_{6}
~+~a_{3}^{\dagger}a_{4}a_{5}~+~H.c.) 
~+~\eta_{7}~\displaystyle \sum_{i \neq j}~ 
a_{i}^{\dagger}a_{i}a_{j}^{\dagger}a_{j}   \\
&~~~+~\eta_{8}~\displaystyle \sum_{i \neq j}~ 
(a_{i}^{\dagger}a_{i}~+~a_{j}^{\dagger}a_{j})(a_{i}^{\dagger}a_{j}~+~
a_{j}^{\dagger}a_{i})  
~+~\eta_{9}~\displaystyle \sum_{i \neq j \neq k}~ 
a_{i}^{\dagger}a_{i}(a_{j}^{\dagger}a_{k}~+~a_{k}^{\dagger}a_{j}) \\
&~~~+~\eta_{10}~\displaystyle \sum_{i \neq j}~ 
(a_{i}^{\dagger}a_{i}^{\dagger}a_{j}a_{j}~+~H.c.) 
~+~\eta_{11}~\displaystyle \sum_{i \neq j \neq k}~ 
(a_{i}^{\dagger}a_{i}^{\dagger}a_{j}a_{k}~+~H.c.) 
\end{array} \eqno (1.5) $$

\noindent
where $\omega$ and $x$ are again used for comparison with
the constants in the Morse potential. $\eta_1$ and $\eta_2$ is the
harmonic constant for the stretch and the bend, respectively.
These $\eta_p$ ($3\leq p \leq 6$) are the Fermi resonance coupling constants,  
$\eta_q$ ($7\leq q \leq 11$) anharmonic constants of the stretching
vibrations, and $\eta_r$ ($12\leq r \leq 15$) anharmonic constants of 
interactions between the stretch and the bend. In Eq. (1.5), we have 
omitted the interaction terms for the bending vibrations, which are similar
to the terms with $\eta_q$ ($7\leq q \leq 11$) for the stretching
modes, and also omitted the fourth-order interactions between the stretching 
and the bending vibrations. In the limit of 
N$_s$ $\rightarrow \infty$ and N$_b$ 
$\rightarrow \infty$, equation (1.5) reduces to the extended local 
mode model for three equivalent stretching bonds $^{[12]}$, where 
the bending vibrations were neglected.

\vspace{4mm}
\begin{center}
{\large \bf 2. Applications}
\end{center}

\noindent
{\bf 2.1 Application to H$_2$O}

\vspace{2mm}
Studies of the vibrations of the water molecule already exist in the
literature. Its stretching vibrational energy levels were calculated
via nonlinear quantum theory $^{[13]}$. Halonen and Carrington 
proposed a simple
vibrational curvilinear internal coordinate hamiltonian $^{[8]}$ 
for its vibrational levels up to 18,500 cm$^{-1}$. The same levels 
were studied by U(4) algebraic model $^{[1]}$. Its recently observed 
vibrational spectrum up to 22,000 cm$^{-1}$ were analyzed by the 
potential-energy surface from {\it ab initio} calculation $^{[14]}$ as well 
as in the boson-realization model and the corresponding $q$-deformed 
model $^{[10]}$.
Owing to much research on this molecule, it provides 
a good testing ground for different models. 

Now, we calculate the Hamiltonian matrix elements in the  
symmetrized bases, then determine the 15 parameters in Eq. (1.4)
by a least-squares optimization in fitting the experimental data 
$E_{obs}$ in cm$^{-1}$ (their normal labels ($\nu_1\nu_2\nu_3$)),
taken from the compilation of Ref. [14].
Two boson numbers N$_s$ and N$_b$ are taken to be 48 and 65, respectively.
The parameters obtained are given in cm$^{-1}$ as follows:
$$\begin{array}{lllll}
\omega_{s}=3705.121, & x_{s}=-4.602, & \omega_{b}=1599.485, & x_{b}=3.913,
& \lambda_{1}=-51.335, \\
\lambda_{2}=-12.746, & \lambda_{3}=-1.004, & \lambda_{4}=3.199, 
& \lambda_{5}=-2.853, & \lambda_{6}=-21.956, \\
\lambda_{7}=12.686, &\lambda_{8}=-0.422,
&\lambda_{9}=-5.901, &\lambda_{10}=2.187, &\lambda_{11}=4.220.
\end{array}  $$

\begin{center}

\fbox{Table 1}

\end{center}

The observed and calculated vibrational levels are presented 
in Table 1 where our results are also compared with those  
calculated in the boson-realization model $^{[10]}$. 
The $\Delta E_{cal}$ are the differences 
between the observed and the calculated values. The 
standard deviation (SD) of this fit is 6.724 cm$^{-1}$, while that of the 
boson-realization model and the corresponding $q$-deformed model 
is 8.198 cm$^{-1}$ and 7.157 cm$^{-1}$ $^{[10]}$, respectively. The present 
model applied to H$_2$O gives a slight improvement in the fit to its 
vibrational energy levels.

\vspace{2mm}
\noindent
{\bf 2.2 Application to AsH$_3$}

\vspace{2mm}
To our knowledge, vibrations of a molecule XY$_3$
have not been studied via algebraic methods yet.
As an example, we now apply Eq. (1.5) to the vibrational spectrum
of arsine (AsH$_3$). The calculation of energy levels will be 
greatly simplified if the symmetrized bases are used. For the 
considered system, these bases can be obtained by the method 
of Ref. [7]. This method is particularly useful when describing 
vibrations of polyatomic molecules and for high overtones. 

Recently observed data ($E_{obs}$) of AsH$_3$
with their normal labels ($\nu_1\nu_2\nu_3^{l_3}\nu_4^{l_4}$)
are taken from the compilation of Refs. [9,15].
Since there are 18 experimental data for this molecule it is not 
possible to determine all coupling constants in Eq. (1.5) by a 
least-squares optimization. It is natural to choose the following
parameters in Eq. (1.5) nonvanishing: $\omega_s$, $x_s$, $\omega_b$, $x_b$, 
$\eta_1$, $\eta_2$, $\eta_3$, $\eta_4$, $\eta_5$, and $\eta_6$.
In other words, we neglect the terms with $\eta_t$ ($7\leq t \leq 11$).
We have found out that other choice does not provide
improvement in our fits. For comparison, we have made
the first fit (Fit a) where the model is in the limit 
of $N_s \rightarrow \infty$ and $N_b \rightarrow \infty$.
The obtained parameters and the
calculated differences ($\Delta E_{cal}^a$) with the corresponding 
observed data ($E_{obs}$) in this fit are listed in Table 2 and 3. 
In another fit (Fit b) we take the boson numbers $N_s$ and $N_b$ to be
62 and 42, respectively. The results in Fit b are also listed in 
the related Tables.

\begin{center}

\fbox{Table 2}

\fbox{Table 3}

\end{center}

The SD in Fit a and b is 10.201 cm$^{-1}$ and 5.827 cm$^{-1}$, respectively.
The calculated results for this molecule in Fit b can be compared 
with those by Lukka {\it et al} $^{[9]}$. However, their vibrational
Hamiltonian was expressed in terms of curvilinear internal coordinates,
and the Morse oscillator basis functions and the results of $ab$ $initio$
calculation were used there.

\vspace{5mm}
\begin{center}
{\large \bf 3. Conclusion and discussion}
\end{center}

We have proposed Fermi resonance-algebraic models for the vibrations
of a bent molecule XY$_2$ and a molecule XY$_3$ by U(2) algebras for the 
description of both stretching and bending modes, where Fermi resonances
between the stretch and the bend are considered.
In the limit, the models reduce to the recently presented boson-realization 
models. As an example, they have been applied to vibrational spectra of the
water molecule and arsine (AsH$_3$). To some extend, they have provided 
improvement in the standard deviations in the corresponding 
boson-realization models for the same observed energy levels. 
It needs more sample calculations to judge which approach is better.

The present calculations and others $^{[2\sim7,10,11]}$ demonstrate
that an algebraic approach can be used as an effective method for
describing molecular vibrations with good precision. In this approach the
eigenvalues and the related wave functions are obtained through
matrix diagonalization. Hence, the required computing time is short,
and some other physical properties, such as transition intensities 
$^{[5,11]}$, are allowed to calculate. It provides the possibility
of using an algebraic model as a numerically efficient, phenomenological
theory for studying molecular spectra, especially when no
{\it ab initio} calculation is available.


\newpage
\begin{center}
{TABLE 1.
Observed and calculated vibrational 
energy levels of $H_2O$ ( $cm^{-1}$) }

\vspace{1mm}
{\small
\begin{tabular}{clrrclrr} 
\hline \hline
($\nu_1\nu_2\nu_3$) & $E_{obs}^{[14]}$  & $\Delta E_{cal}^{[10]}$ & 
 $\Delta E_{cal}$ & 
($\nu_1\nu_2\nu_3$) & $E_{obs}^{[14]}$  & $\Delta E_{cal}^{[10]}$ & 
 $\Delta E_{cal}$ \\
\hline
(010) & 1594.7498 & $-2.668$ & $-$4.733 & 
~(001)& 3755.93   &   0.170 & $-$0.522\\[-2mm]
(020)& 3151.63   &   0.016  & $-$4.136 &
~(011)& 5331.269  &     $-5.434$ & $-$5.562\\[-2mm]
(100)& 3657.053  &   3.549 & 1.997 &
~(021)& 6871.51   &     $-3.398$ & $-$4.325\\[-2mm]
(030)& 4666.793  &  5.464  & 0.110 &
~(101)& 7249.81   &    2.611    & 0.409\\[-2mm]
(110)& 5234.977  & 1.684   & 0.694 &
~(031)& 8373.853  &    4.051 & 2.443\\[-2mm]
(040)& 6134.03   &  9.088  & 4.828 &
~(111)& 8807      &     $-1.858$  & $-$0.673 \\[-2mm]
(120)& 6775.1    &  1.817  & $-$0.591& 
~(041)& 9833.584  & 13.643   & 12.564 \\[-2mm]
(200)& 7201.54   &    1.302& 0.483 &
~(121)& 10328.731 &     $-0.616$   & $-$0.006\\[-2mm]
(002)& 7445.07   & 3.063   & 1.453 &
~(201)& 10613.355 &   0.766   & $-$0.809\\[-2mm]
(050)& 7552      &  11.254 & 10.415&
~(003)& 11032.406 &    5.575  & 3.577 \\[-2mm]
(130)& 8273.976  &  0.594  & $-$1.912& 
~(131)& 11813.19  &    2.636  & 1.919\\[-2mm]
(210)& 8761.582  &   0.248 & 1.828 &
~(211)& 12151.26  &   $-3.666$& 0.371 \\[-2mm]
(060)& 8890      & $-17.089$& $-$12.506 &
~(013)& 12565     &     $-9.461$& $-$6.151\\[-2mm]
(012)& 9000.136  & $- 5.600$& $-$4.650 &
~(141)& 13256     & 4.292    & 4.464\\[-2mm]
(220)& 10284.367 &   $- 0.405$& $-$0.714& 
~(221)& 13652.656 &   $-3.450$  & $-$0.116 \\[-2mm]
(022)& 10524.3   &  $-0.950 $& $-$0.991&
~(301)& 13830.938 &   0.384  & $-$1.448\\[-2mm]
(300)& 10599.686 &   $-3.117$& $-$3.002&
~(023)& 14066.194 &     $-9.806$   & $-$7.824 \\[-2mm]
(102)& 10868.876 &  5.583 & 1.810&
~(103)& 14318.813 &   6.331  & 1.985\\[-2mm]
(310)& 12139.2   &  $-6.313$ & $-$1.508 &
~(151)& 14640     & $-10.977$   & $-$5.968 \\[-2mm]
(112)& 12407.64  &  3.124   & 3.469 &
~(231)& 15119.029 &   $-3.536$  & $-$2.852 \\[-2mm]
(042)& 13448     &    7.777 & 4.483 &
~(311)& 15347.956 &   $-4.276$  & 1.217 \\[-2mm]
(320)& 13642.202 &   $-7.521$& $-$4.024 &
~(113)& 15832.765 &   $-7.239$  & $-$1.381 \\[-2mm]
(202)& 13828.277 &   2.052 & 0.344 &
~(321)& 16821.635 &   $-2.039$  & 2.715\\[-2mm]
(122)& 13910.896 &   6.188 & $-$0.130 &
~(203)& 16898.842 &   3.234 & $-$1.570 \\[-2mm]
(400)& 14221.161 &     $-1.522$ & $-$1.806 &
~(123)& 17312.539 &   1.834 & $-$1.041\\[-2mm]
(004)& 14536.87  &    9.844 & 6.393 &
~(401)& 17495.528 &   0.756 & $-$0.563 \\[-2mm]
(330)& 15107     &   $-14.579$ & $-$12.251 &
~(331)& 18265.82  & $-3.654$ & $-$2.516\\[-2mm]
(212)& 15344.503 &   12.518  & 10.038 &
~(213)& 18393.314 &   $-1.400$& 0.480\\[-2mm]
(410)& 15742.795 &   0.179   & 4.839 &
~(411)& 18989.961 &   $-12.986$ & $-$0.621\\[-2mm]
(222)& 16825.23  &   $-16.629$& $-$18.041 &
~(421)& 19720     & $-0.801$  & $-$0.492\\[-2mm]
(302)& 16898.4   &  1.469   & $-$3.711 &
~(303)& 19781.105 & 14.596  & 8.509 \\[-2mm]
(420)& 17227.7   & 11.514   & 0.715 &
~(501)& 20543.137 & $-7.555$ & $-$2.973\\[-2mm]
(500)& 17458.354 &  0.824   & 0.139 &
~(313)& 21221.828 & $-6.455$ & $-$5.430\\[-2mm] 
(104)& 17748.073 &     $-0.279$ & $-$0.504 &
(232)& 18320     &  25.584  & 29.589 \\ [-2mm]
(312)& 18392.974 &    $-1.070$ & 0.979 &
(412)& 21221.569 &  $-5.258$   & $-$5.451 \\
\cline{5-8}
& & & & & SD     & 8.198      & 6.724\\
\hline
\hline
\end{tabular}  }
\end{center}

\newpage
~~
\vspace{8mm}

\begin{center}
{TABLE 2.
Parameters (cm$^{-1}$) for AsH$_3$ obtained by the least 
square fitting }

\vspace{2mm}
\begin{tabular}{cccccc} 
\hline
\hline
&{$\omega_s$} &{$x_s$ } &{$\omega_b$} &{$x_b$ } & {$\eta_1$} \\
Fit a & 2039.921 & $-$80.580 & 959.021 & 64.320 & $-$24.775 \\
Fit b & 2066.972 & $-$28.376 & 962.647 & 68.865 & $-$18.639 \\
\hline
&&&&&\\
\hline
& {$\eta_2$} & {$\eta_3$} & {$\eta_4$} & {$\eta_5$} & {$\eta_6$} \\
Fit a & $-$21.932
& $-$53.616 & $-$5.996 & $-$33.712 & 1.823 \\
Fit b & $-$27.283
& $-$55.482 & $-$5.850 & $-$29.862 & 11.970 \\
\hline  \hline  
\end{tabular}  
\end{center}

\vspace{20mm}

\begin{center}
{TABLE 3.
Observed and calculated vibrational 
energy levels of AsH$_3$ (cm$^{-1}$) }

\vspace{1mm}
{\small
\begin{tabular}{clrrclrr} 
\hline \hline
($\nu_1\nu_2\nu_3^{l_3}\nu_4^{l_4}$) & $E_{obs}^{[9,15]}$ &$\Delta E_{cal}^a$
& $\Delta E_{cal}^b$ & 
($\nu_1\nu_2\nu_3^{l_3}\nu_4^{l_4}$) & $E_{obs}^{[9,15]}$ &$\Delta E_{cal}^a$
& $\Delta E_{cal}^b$ \\
\hline
($010^00^0$) & 906.752 & $-$8.425 & $-$1.328 & 
($000^01^{\pm1}$) & 999.225 & 8.253 & 9.295 \\

($020^00^0$) & 1806.149 & 3.950 & 1.853 & 
($010^01^{\pm1}$) & 1904.115 & $-$9.368 & $-$0.427 \\

($000^02^0$) & 1990.998 & $-$2.224 & 1.602 & 
($000^02^{\pm2}$) & 2003.483 & $-$10.736 & $-$9.707 \\

($100^00^0$) & 2115.164 & $-$0.720 & 0.932 & 
($001^10^0$) & 2126.423 & 2.698 & $-$1.537 \\

($110^00^0$) & 3013 & 1.704 & 3.953 & 
($001^11^{\pm1}$) & 3102 & $-$0.615 & 3.396 \\

($200^00^0$) & 4166.772 & $-$0.182 & 3.033 & 
($101^10^0$) & 4167.935 & $-$8.389 & $-$2.245 \\

($300^00^0$) & 6136.316 & 2.401 & $-$4.667 & 
($201^10^0$) & 6136.310 & 9.419 & 1.951 \\

($003^00^0$) & 6276 & $-$3.991 & $-$2.321 & 
($102^20^0$) & 6295 & 2.410 & 2.238 \\

($400^00^0$) & 8028.977 & $-$2.275 & $-$1.389 & 
($301^10^0$) & 8028.969 & 1.398 & 0.713 \\
\cline{5-8}
& & & & &SD     & 10.201      & 5.827\\
\hline
\hline
\end{tabular}  }
\end{center}

\end{document}